\documentclass[conference]{IEEEtran}
\IEEEoverridecommandlockouts
\usepackage{cite}
\usepackage{booktabs}
\usepackage{multirow}

\usepackage{amsmath,amssymb,amsfonts}
\usepackage{algorithmic}
\usepackage{graphicx}
\usepackage{subfigure}
\usepackage{textcomp}
\usepackage{xcolor}
\def\BibTeX{{\rm B\kern-.05em{\sc i\kern-.025em b}\kern-.08em
    T\kern-.1667em\lower.7ex\hbox{E}\kern-.125emX}}
\begin{document}

\title{Asymmetric Adaptation-based Real-time Fault Diagnosis Under Transitional Operating Conditions
}

\author{\IEEEauthorblockN{Hongshuo Zhao}
\IEEEauthorblockA{
\textit{MCC5 Group Shanghai Co. LTD}\\
Shanghai, China \\
zhs17603218096@163.com}
\and
\IEEEauthorblockN{Zeyi Liu}
\IEEEauthorblockA{\textit{Tsinghua University} \\
Beijing, China \\
liuzy21@mails.tsinghua.edu.cn}
\and
\IEEEauthorblockN{Xiao He}
\IEEEauthorblockA{\textit{Tsinghua University} \\
Beijing, China \\
hexiao@tsinghua.edu.cn}

% \and
% \and
% \and
}

\maketitle

\begin{abstract}
Data streams in real-world industrial scenarios often contain transitional operating conditions that are uncovered during offline training, leading to significant distribution shifts. To bridge the gap between static offline models and dynamic online data, a novel asymmetric adaptation-based fault diagnosis method is proposed in this paper. Specifically, in the offline stage, we employ domain generalization techniques to extract domain-invariant features from multiple stable conditions and construct robust normalized fault prototypes as reference anchors. Subsequently, during online inference, we design an online test-time adaptation method based on a periodic prototype re-projection mechanism to dynamically update prototype positions. Furthermore, we utilize the geometric distribution derived from anchors to guide the updates of classifiers and adopt an asymmetric learning rate strategy for the feature extractor and classifier. The proposed approach ensures rapid adaptation to new transitional conditions while preserving the discriminative power inherited from the offline domain generalization initialization. Experimental results demonstrate that this mechanism effectively leverages offline generalized knowledge to guide online inference, significantly improving robustness in non-stationary environments.
\end{abstract}


\section{Introduction}

\IEEEPARstart{D}{ata}-driven fault diagnosis has emerged as a pivotal technology for ensuring the reliability and safety of complex industrial assets \cite{10292791,henao2014trends,gangsar2022review,fink2020potential}. With the development of advanced sensors and the accumulation of massive monitoring data, deep learning algorithms have demonstrated remarkable proficiency in extracting fault features. Under stationary operating conditions where training and testing data share identical statistical distributions, these methods have achieved near-perfect diagnostic accuracy \cite{liu2018artificial}.

However, the assumption of {\it independent and identically distributed} (i.i.d.) data is rarely satisfied in practical industrial scenarios \cite{11104131,sun2024continuous,9869794}. Complex machinery often operates under dynamic conditions, involving frequent fluctuations in load, speed, and environmental factors. Such variations introduce significant distribution discrepancies, widely known as domain shifts, between the source domain (offline training data) and the target domain (online testing data) \cite{10342662,sammaknejad2015operating}. Although {\it domain generalization} (DG) has been extensively studied to mitigate such discrepancies, DG methods typically necessitate access to a comprehensive set of target domain data during the training phase \cite{zhao2022domain,fan2023deep,guo2024domain,guo2024chemical}. Such a requirement is impractical for real-time monitoring systems where data arrives in sequential streams with time-varying operating conditions.

A more severe challenge arises from \textit{transitional operating conditions} (e.g., run-up or run-down processes) \cite{borucki2012diagnosis,11007216,wang2019domain,peng2017multimode}. Unlike discrete stationary states, transitional conditions induce continuous and non-stationary distribution drifts driven by time-varying parameters. In such dynamic environments, static models trained solely on stable conditions suffer from rapid performance degradation. To address this difficulty, \textit{test-time adaptation} (TTA) has attracted increasing attention by enabling model adaptation on the test stream without accessing source data \cite{boudiaf2022parameter,wang2025search,liang2025comprehensive}. The prevailing TTA paradigm relies on self-training via pseudo-labeling, utilizing high-confidence predictions as supervision signals. Nevertheless, under severe domain shifts caused by transitional regimes, the initial classifier often becomes unreliable and generates biased predictions, termed \textit{semantic noise}. Directly retraining the model with such noisy labels inevitably leads to error accumulation and the catastrophic failure of the diagnostic system.

In this paper, a novel framework is proposed to overcome the aforementioned limitations. The core motivation is to bridge the gap between offline static knowledge and online dynamic stream. Specifically, DG techniques are employed in the offline stage to extract domain-invariant features and construct normalized fault prototypes. Subsequently, in the online phase, an adaptation mechanism is introduced. The main contributions of this work are summarized as follows:

\begin{itemize}
    \item A novel diagnosis framework is established, which integrates offline DG with online TTA to tackle the challenge of unseen transitional operating conditions.
    
    \item An asymmetric update strategy is developed, which makes the model tracks rapid distribution changes in the online stream while preserving the robustness of the latent feature space.
    
    \item Numerous experiments are constructed based on the opern-sourced MCC5-THU Gearbox datasets. The results demonstrate the effectiveness of the proposed scheme in coping with transitional operating conditions.

\end{itemize}

The remainder of this paper is organized as follows.
 Section \ref{section_main_procedure} describes the proposed method in detail, including both the offline initialization and online adaptation stages. Section \ref{section_experiment} reports the experimental setup and discusses the corresponding results. Finally, Section \ref{section_conclusion} summarizes the main findings and conclusions of the study.

\section{The Proposed Approach}\label{section_main_procedure}

\subsection{Problem Formulation and Notations}
Consider an industrial fault diagnosis task involving $K$ distinct fault categories. In the initial phase, a small-scale, clean, and labeled dataset $\mathcal{D}^{(0)}=\{(x_i^{(0)},y_i^{(0)},c_i^{(0)})\}_{i=1}^{N^{(0)}}$ is acquired under stationary operating conditions for offline initialization. Here, $x_i^{(0)}\in\mathbb{R}^{d}$ represents the monitoring sample extracted from windowed multi-channel vibration signals, $y_i^{(0)}\in\{1,2,\dots,K\}$ denotes the corresponding fault category label, and $c_i^{(0)}\in\{1,2,\dots,M\}$ indicates the operating condition index associated with $x_i^{(0)}$. An offline diagnosis model $f_{\theta}$ is constructed to obtain the initialized parameters $\theta^{(0)}$.

In the online phase, data arrives in a sequential stream $\mathcal{S}=\{\mathcal{B}^{(1)},\mathcal{B}^{(2)},\ldots\}$, where each data block $\mathcal{B}^{(t)}=\{x_i^{(t)}\}_{i=1}^{N^{(t)}}$ contains $N^{(t)}$ unlabeled monitoring samples $x_i^{(t)}\in\mathbb{R}^{d}$ collected during operation. In practical deployment scenarios, neither ground-truth fault labels $y_i^{(t)}$ nor explicit operating condition information is assumed to be accessible at test time.

A critical challenge arises from the fact that the online data stream may contain \textit{transitional operating conditions} not covered during offline training. Such a phenomenon results in significant distribution shifts relative to the stationary initialization data. The objective is to mitigate the degradation of diagnostic performance under the aforementioned transitional conditions.

\subsection{Offline Stage}

The primary objective of the offline stage is to establish a robust baseline diagnosis model utilizing historical data collected under stationary conditions. For the raw multi-channel system measurements from the source domain, a fixed-length sliding window technique is first employed to segment continuous time-series data into discrete monitoring samples $x \in \mathbb{R}^d$. The described preprocessing step constructs the labeled source domain dataset $\mathcal{D}^{(0)}$, which encapsulates the feature characteristics of $K$ fault types under $M$ different stationary operating conditions.

Since fault features are highly sensitive to dynamic operating parameters (such as varying speeds and loads), the direct application of a standard classifier may result in performance deterioration. To alleviate the impact of such distribution changes and enhance the generalization capability against domain shifts, a {\it domain adversarial neural network} (DANN) is adopted \cite{ganin2016domain,sicilia2023domain}. The framework comprises a feature extractor $G_f(\cdot; \theta_f)$, a fault classifier $G_y(\cdot; \theta_y)$, and an auxiliary condition discriminator $G_d(\cdot; \theta_d)$. By treating distinct stationary conditions as separate domains, the discriminator aims to infer the condition index $c$ from the features, while the extractor is trained to confuse the discriminator via a min-max adversarial game. The optimization objective is formalized as follows:
\begin{equation}
\begin{aligned}
\mathcal{L}_{\text{offline}} & = \mathbb{E}_{(x, y) \sim \mathcal{D}^{(0)}} \left[ \mathcal{L}_{\text{ce}}(G_y(G_f(x)), y) \right] \\
& - \lambda \mathbb{E}_{(x, c) \sim \mathcal{D}^{(0)}} \left[ \mathcal{L}_{\text{ce}}(G_d(G_f(x)), c) \right],
\end{aligned}
\end{equation}
where $\mathcal{L}_{\text{ce}}$ denotes the cross-entropy loss, and $\lambda$ balances diagnostic accuracy and condition invariance. Upon convergence, the discriminator $G_d$ is discarded, and the parameters $\theta^{(0)} = \{\theta_f, \theta_y\}$ are retained, yielding a feature space that is not sensitive to condition variations.

To further facilitate robust adaptation, a static anchor memory bank $\mathcal{M}_{\text{off}}$ is constructed to preserve reliable fault patterns learned from the stable data. A prototype $\mu_k$ is computed for each fault category $k$ as the geometric center in the feature space. Considering that signal energy fluctuates under varying conditions, normalized feature representations are leveraged to focus on the directional alignment of fault signatures:
\begin{equation}
\mu_k = \frac{1}{|\mathcal{M}_{\text{off}}^k|} \sum_{x \in \mathcal{M}_{\text{off}}^k} \frac{G_f(x)}{\|G_f(x)\|_2},
\end{equation}
where $\mathcal{M}_{\text{off}}^k$ denotes the set of source domain anchor samples belonging to the $k$-th fault class. Such prototypes $\mathcal{P} = \{\mu_1, \mu_2, \dots, \mu_K\}$ serve as reliable reference anchors for the subsequent online stage.

\subsection{Online Stage}
Although the offline stage attempts to minimize the impact of condition variations on fault distributions by seeking domain-invariant features, theoretical convergence of the pre-trained model is often difficult to guarantee in practice. Furthermore, the specific impact of transitional conditions on fault distributions is typically unavailable in advance. Consequently, further adjustment in the online stage is crucial.

In this paper, a novel adaptation mechanism is proposed for the online stage. The diagnostic system is presented with a sequential data stream $\mathcal{S}=\{\mathcal{B}^{(t)}\}_{t=1}^{\infty}$ collected under varying and transitional conditions. To address the aforementioned difficulty, an online TTA method is introduced to dynamically align feature geometry with semantic predictions via noise transfer modeling.

In the categorical probability space, self-training based on pseudo-labels is a common adaptation strategy in the absence of ground-truth labels. However, due to distribution shifts in transitional conditions, the classifier $G_y$ initialized on stable data inevitably produces biased predictions, referred to as \textit{semantic noise}. Such error patterns are often systematic (e.g., fault A is consistently misidentified as fault B under specific transient loads). Directly utilizing raw predictions from a biased classifier as supervision signals leads to error accumulation. Therefore, modeling the error transfer channel is aimed at mitigating the cumulative impact of such errors.

For each sample $x$, the classifier outputs ${p}_{\mathrm{cls}}(x)=\left[p_{\mathrm{cls}}(1 \mid x), p_{\mathrm{cls}}(2 \mid x), \ldots, p_{\mathrm{cls}}(K \mid x)\right]^{\top}$, where:

\begin{equation}
p_\text{cls}(k \mid x)=\operatorname{softmax}\left(G_y\left(G_f(x)\right)\right)_k.
\end{equation}

Simultaneously, based on the similarity between normalized features and prototypes $s_k(x)=\left\langle\bar{z}, \bar{\mu}_k\right\rangle$, a geometric distribution is defined as:

\begin{equation}
q_{\text {geo }}(k \mid x)=\frac{\exp \left(s_k(x) / \tau\right)}{\sum_{j=1}^K \exp \left(s_j(x) / \tau\right)}.
\end{equation}
where $\tau$ denotes the temperature coefficient.
The geometric pseudo-label and the classifier's hard prediction are defined respectively as:
\begin{equation}
\hat{y}_{\mathrm{geo}}(x)=\arg \max _k q_{\mathrm{geo}}(k \mid x)
\end{equation}
and
\begin{equation}
\quad \hat{y}_{\mathrm{cls}}(x)=\arg \max _k p_{\mathrm{cls}}(k \mid x) .
\end{equation}

% To reduce the cognitive bias introduced by utilizing static memory alone, an online memory bank $\mathcal{M}_{on}$ is maintained to store relatively reliable samples from the recently received system measurements $\mathcal{B}^{(t)}$. The high-confidence set is defined as:
% \begin{equation}
% \Omega_t=\left\{x \in \mathcal{B}^{(t)}: \max _k p_{\text {geo }}(k \mid x) \geq \delta\right\},
% \end{equation}
% where $\delta$ is a preset hyperparameter threshold. The tuples $\left\{\left(x, \hat{y}_{\text {geo }}, p(\cdot \mid x)\right)\right\}_{x \in \Omega_t}$ are stored in $\mathcal{M}_{on}$ using a First-In-First-Out (FIFO) policy. 
To ensure the validity of geometric stability in the latent feature space, the classifier and extractor are jointly updated:
\begin{equation}
\mathcal{L}_{\mathrm{online}}^{(t)}=-\mathbb{E}_{x \sim \mathcal{B}^{(t)}} \sum_{k=1}^K {q}_{\text {geo}}(k \mid x) \log p_{\mathrm{cls}}(k \mid x).
\end{equation}

Notably, to more rapidly track feature drift induced by the extractor, $n$ inner-loop updates are executed after the arrival of each data block, thereby achieving faster online adaptation under limited sample and label-free constraints. Formally, let $(\theta_f^{(t, 0)}, \theta_y^{(t, 0)})=(\theta_f^{(t-1)}, \theta_y^{(t-1)})$. Then, for $r=1, 2, \ldots, n$:
\begin{equation}
\theta_f^{(t, r)} = \theta_f^{(t, r-1)}-\eta_f \nabla_{\theta_f} \mathcal{L}_{\mathrm{online}}^{(t, r-1)}
\end{equation}
and
\begin{equation}
\quad \theta_y^{(t, r)} = \theta_y^{(t, r-1)}-\eta_y \nabla_{\theta_y} \mathcal{L}_{\mathrm{online}}^{(t, r-1)}
\end{equation}
where $\mathcal{L}_{\mathrm{online}}^{(t, r-1)}$ is recalculated on data block $\mathcal{B}^{(t)}$ based on the current inner-loop parameters $\{\theta_f^{(t, r-1)}, \theta_y^{(t, r-1)}\}$. The learning rates are set such that $\eta_f < \eta_y$ to implement an update strategy with consistently differential stability magnitudes.

To accommodate the constantly evolving feature space, the static anchor memory bank $\mathcal{M}_\text{off}$ is utilized to periodically re-project the prototypes. Every $N_{f}$ steps, the prototypes are updated as follows:
\begin{equation}
\mu^{(t)}_k = \frac{1}{|\mathcal{M}_{\text{off}}^k|} \sum_{x \in \mathcal{M}_{\text{off}}^k} \left[\frac{G^{(t)}_f(x)}{\|G^{(t)}_f(x)\|_2}\right].
\end{equation}

% \begin{figure}[htbp]
% \centerline{\includegraphics{fig1.png}}
% \caption{Example of a figure caption.}
% \label{fig}
% \end{figure}

\section{Experiments} \label{section_experiment}

\subsection{Datasets}

\begin{figure}[htbp]
    \centering
    \includegraphics[width=0.49\textwidth]{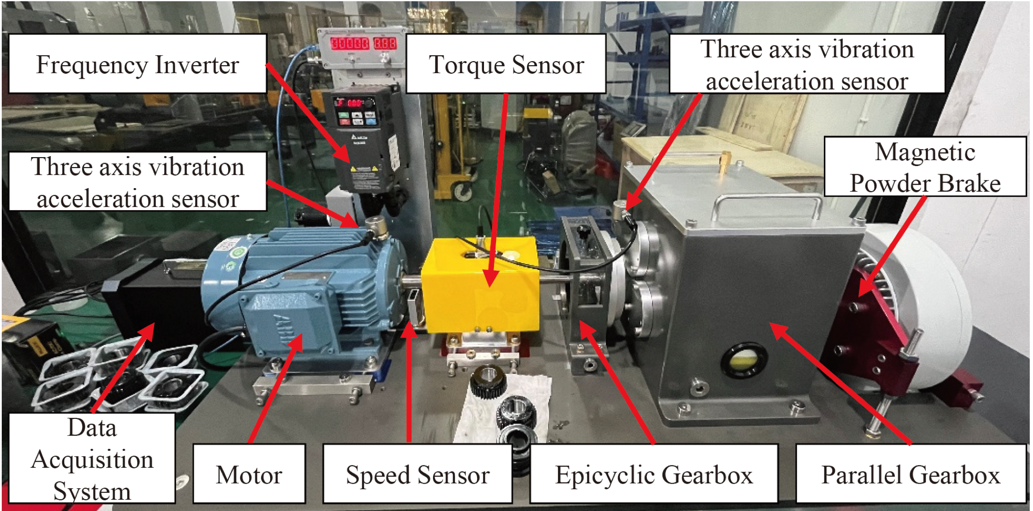}
    \caption{The actual gearbox test rig \cite{chen2024multimode}.}
    \label{fig_equipment}
\end{figure}

In this study, we utilize a dataset collected from a two-stage parallel gearbox test rig to evaluate fault diagnosis performance under time-varying conditions. As shown in Fig. \ref{fig_equipment}, the rig consists of a 2.2 kW three-phase asynchronous motor, a measurement and control system, and a magnetic powder brake that applies load torque. The torque and speed are monitored by an S2001 torque sensor (accuracy: $\pm$ 0.5\% F.S) and a key phase speed sensor, respectively.

\begin{figure}[htbp]
    \centering
    \includegraphics[width=0.45\textwidth]{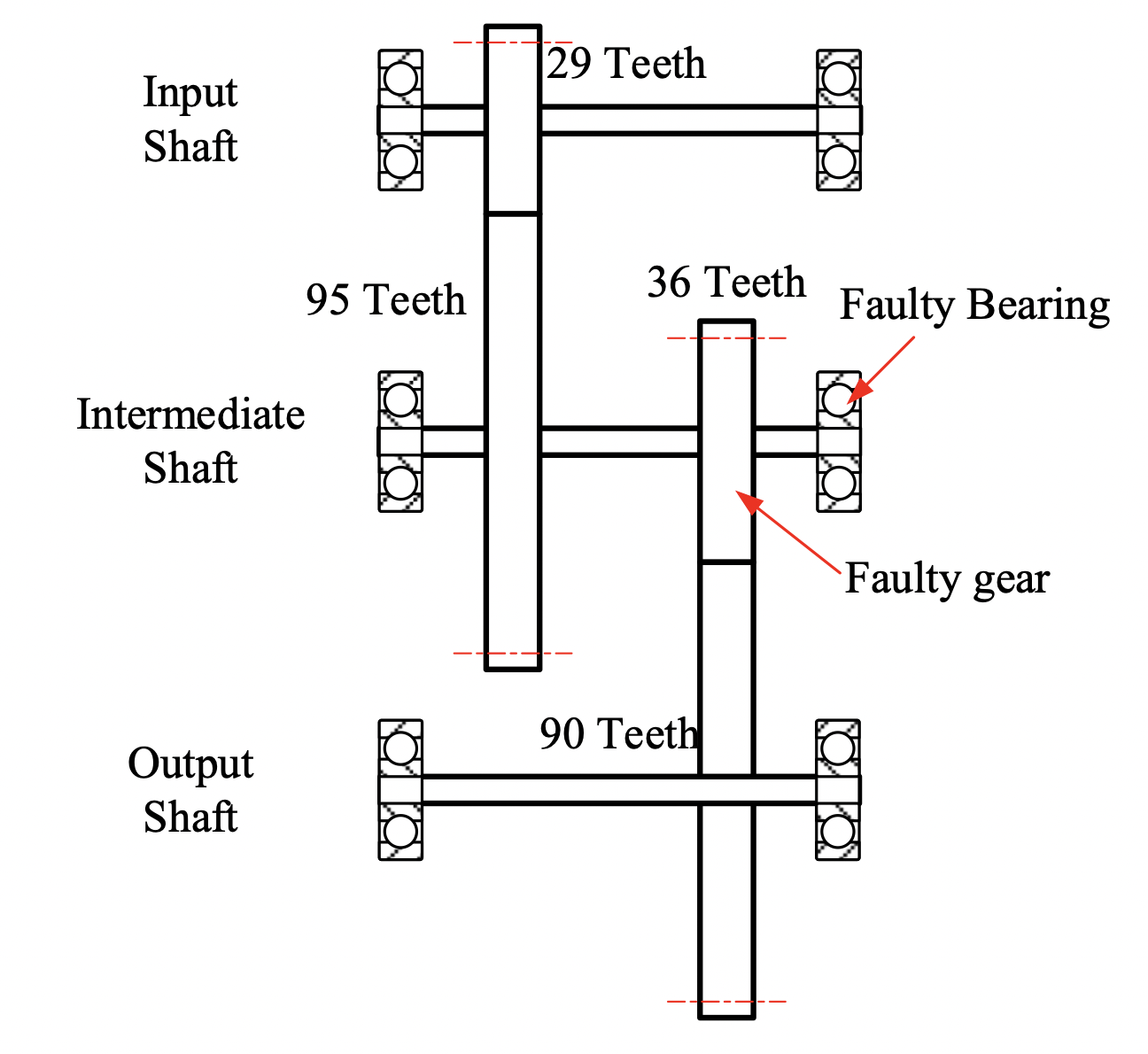}
    \caption{The internal structure diagram of the gearbox \cite{chen2024multimode}.}
    \label{fig_inter}
\end{figure}

For data acquisition, two TES001V three-axis accelerometers (sensitivity: 100 mV/g) were deployed to measure vibration signals from the motor and the gearbox intermediate shaft along the x, y, and z axes. The sampling rate was set to 12.8 kHz. The experiments were conducted under 12 different working conditions with the laboratory temperature variation controlled within 2$^{\circ}$C to mitigate environmental noise.
The dataset contains 240 sets of time-series data covering healthy states, single faults, and compound faults. The specific components of interest are the 36-tooth gear (module 1.5, width 10 mm) on the intermediate shaft and the adjacent ER16K support bearing (see Fig. \ref{fig_inter}). All faults were precisely manufactured using laser etching with 0.01 mm accuracy to simulate varying fault types and degrees.
It encompasses the healthy state, gear wear, teeth crack, teeth break, gear pitting, missing teeth, and two types of compound faults. Detailed descriptions of these fault patterns are reported in \cite{chen2024multimode}.

\begin{table*}[htbp]
  \centering
  \caption{Description of Experimental Working Conditions for Offline and Online Stages.}
  \label{tab:experimental_settings}
  \renewcommand{\arraystretch}{1.2}
  \setlength{\tabcolsep}{1pt} % 可选：增大列间距（默认约6pt）

  \begin{tabular*}{\textwidth}{@{\extracolsep{\fill}}cccccc}
    \toprule
    \textbf{Scenario} & \textbf{Stage} & \textbf{Speed (rpm)} & \textbf{Torque (Nm)} & \textbf{Condition Type} & \textbf{\#Samples} \\
    \midrule

    \multirow{6}{*}{\textbf{Scenario I}}
      & \multirow{3}{*}{\textbf{Offline}}
      & 1000 & 10 & Steady & 6344 \\
      & & 1000 & 15 & Steady & 6344 \\
      & & 1000 & 20 & Steady & 6344 \\
      \cmidrule{2-6}
      & \multirow{3}{*}{\textbf{Online}}
      & 1000 & 20 & Steady & 7143 \\
      & & 1000 & $20 \to 15$ & \textbf{Transitional} & 3936 \\
      & & 1000 & 15 & Steady & 7143 \\
    \midrule

    \multirow{6}{*}{\textbf{Scenario II}}
      & \multirow{3}{*}{\textbf{Offline}}
      & $1000$ & 20 & Steady & 6344 \\
      & & $1500$ & 20 & Steady & 6344 \\
      & & $2000$ & 20 & Steady & 6344 \\
      \cmidrule{2-6}
      & \multirow{3}{*}{\textbf{Online}}
      & $2000$ & 20 & Steady & 6344 \\
      & & $2000  \to 1500$ & 20 & \textbf{Transitional} & 6344 \\
      & & $1500$ & 20 & Steady & 6344 \\

    \bottomrule
  \end{tabular*}
\end{table*}

\subsection{Experimental Settings}

\begin{figure*}[htbp]
     \centering
     \subfigure[Scenario I, Gear pitting]{
               \includegraphics[width=0.45\textwidth]{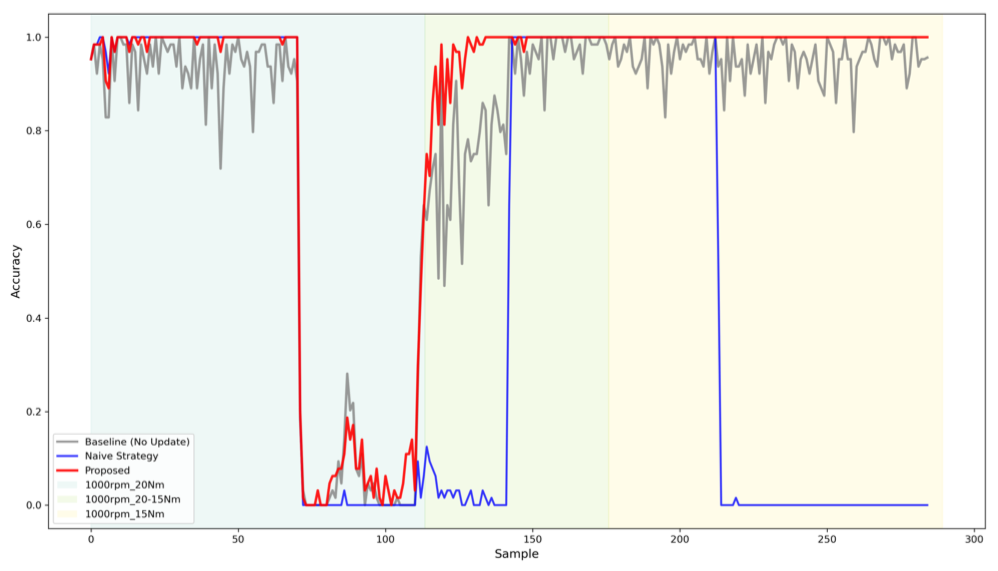}
     }
     \subfigure[Scenario II, Gear pitting]{
               \includegraphics[width=0.45\textwidth]{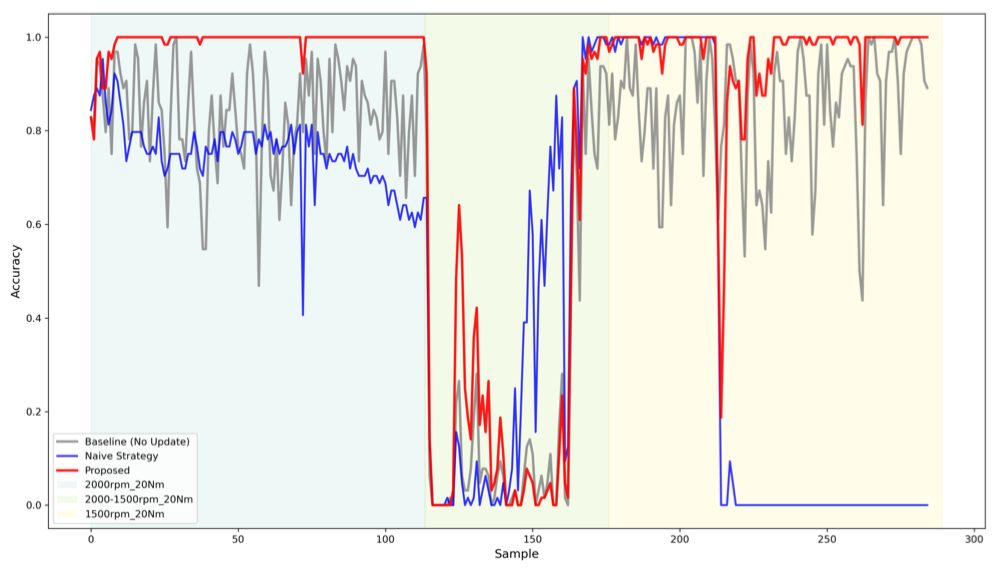}
     }
     \subfigure[Scenario I, Gear wear]{
               \includegraphics[width=0.45\textwidth]{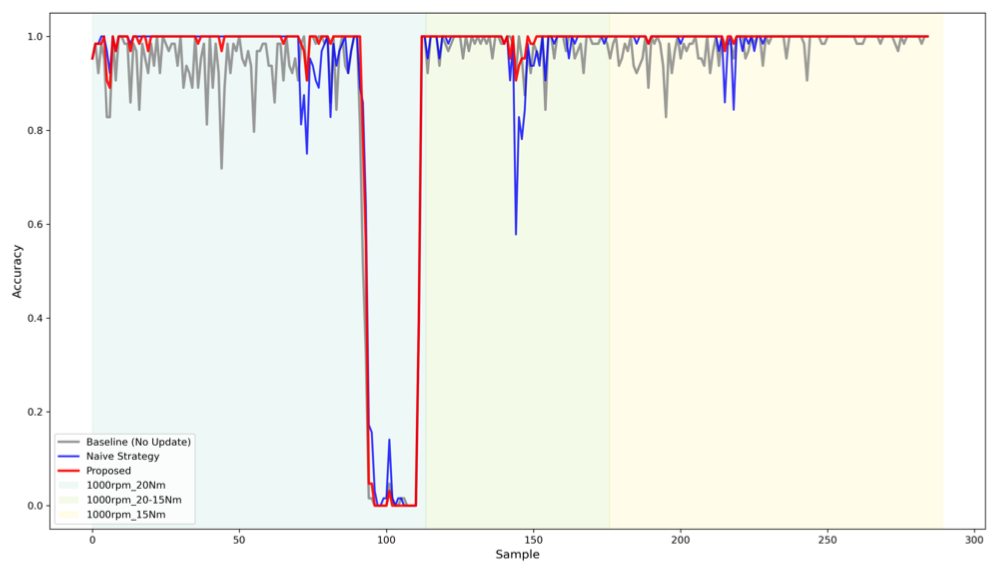}
     }
     \subfigure[Scenario II, Gear wear]{
               \includegraphics[width=0.45\textwidth]{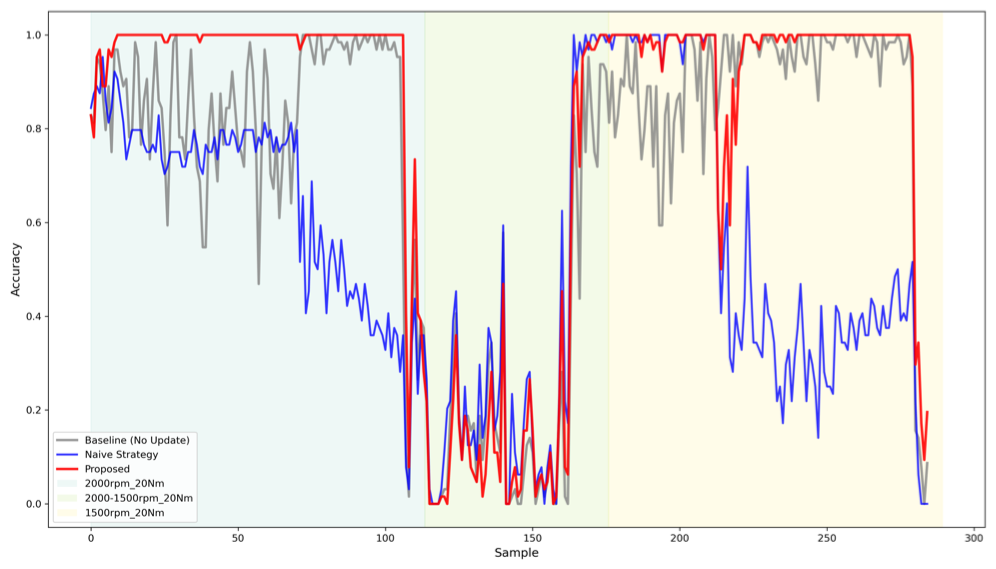}
     }
     \subfigure[Scenario I, Teeth crack]{
               \includegraphics[width=0.45\textwidth]{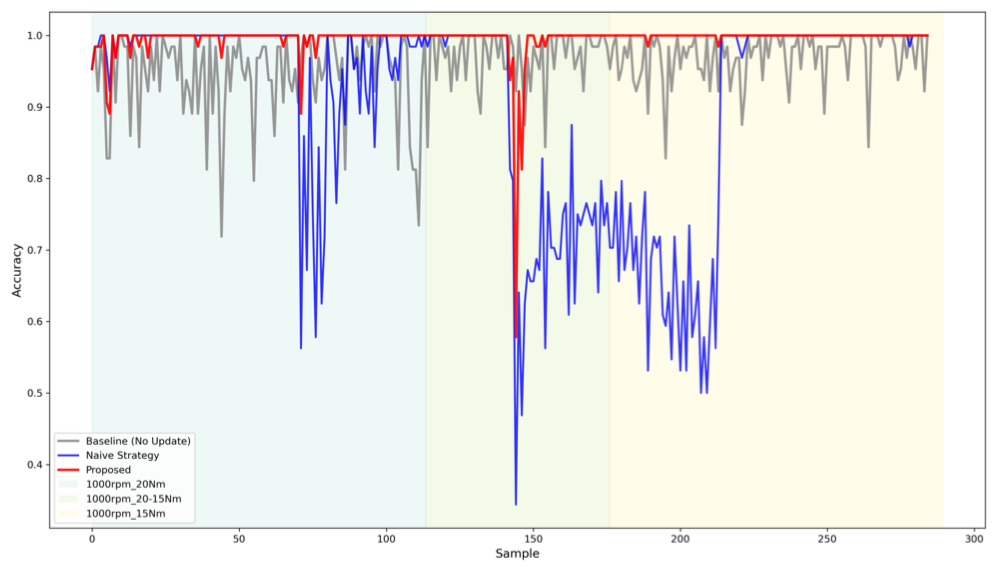}
     }
     \subfigure[Scenario II, Teeth crack]{
               \includegraphics[width=0.45\textwidth]{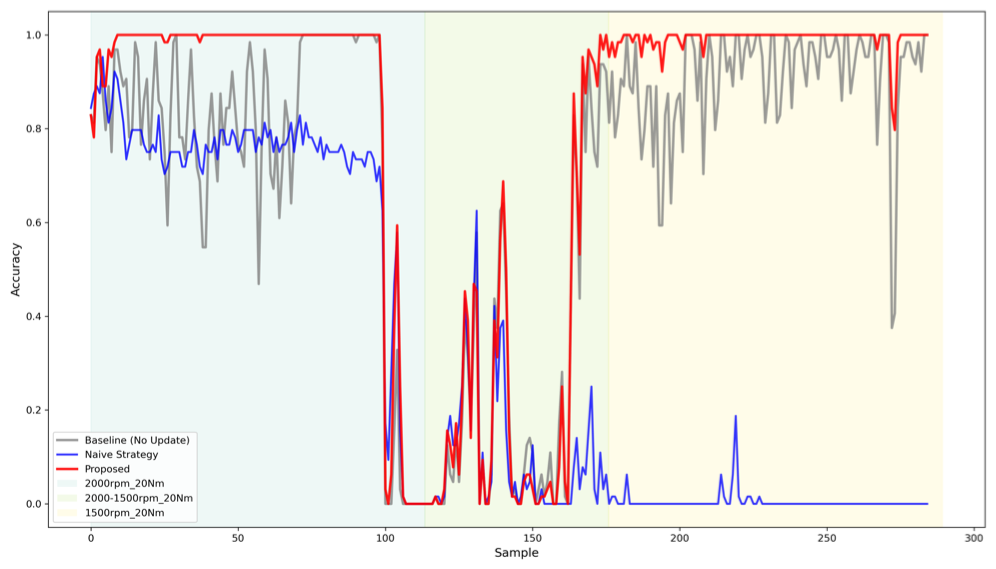}
     }
     \caption{The results of real-time performance in the comparison study.}
     \label{Exp}
\end{figure*}

A sliding-window strategy with a length of 1024 points and a stride of 16 points is employed to segment the raw multi-channel vibration recordings, yielding highly overlapped time-series segments. For each sample, signals from the six vibration channels are concatenated to form the input tensor for the model. Fault labels are derived from the original recordings and re-mapped to consecutive class indices based on a predefined fault set.

In the offline training phase, the source-domain training set is constructed using data collected under three steady operating conditions with varying load torques and varying speed. The detailed information is reported as Table \ref{tab:experimental_settings}. To prevent data leakage while preserving the temporal characteristics of the recordings, only the initial 10\% of the windowed samples for each fault category under each training condition are utilized for offline model initialization.

To evaluate diagnostic performance under time-varying conditions, a continuous test stream incorporating the target-domain sequence is constructed. The stream comprises the remaining 90\% of the samples from the steady conditions (held out during training) and all samples acquired under transitional conditions. To simulate the intermittent occurrence and recurrence of faults during operating condition changes, target fault samples are introduced into the stream at approximately the 25\% timestamp, reverted to the healthy state at the 50\% timestamp, and re-introduced at the 75\% timestamp. 

\subsection{Architecture}

In the offline training stage, a DANN architecture was adopted, and three main components were included: a feature extractor, a fault classifier, and a domain discriminator. As input to the feature extractor, a flattened multi channel vibration signal was used. Each sample was formed by concatenating six vibration channels within a sliding window of 1024 data points, including three axes from an accelerometer on the motor side and three axes from an accelerometer on the gearbox intermediate shaft. As a result, the input dimension was 6144. The feature extractor was constructed using four fully connected layers with progressively reduced dimensions, where the layer widths were set to 6144, 1024, 512, 256, and 64. Batch normalization and a ReLU activation function were applied after each layer. The resulting 64 dimensional feature representation was then fed into a two layer classifier network with dimensions of 64, 32, and $K$, where $K$ denotes the number of fault categories. A ReLU activation function was applied in the hidden layer, and the output logits were produced for fault diagnosis.

To support domain invariant feature learning across multiple source domains, a domain discriminator was introduced. The discriminator was built using four fully connected layers with dimensions of 64, 128, 128, 64, and $M$, where $M$ denotes the number of source domains. Batch normalization was applied in the third layer. During offline training, a gradient reversal layer was placed between the feature extractor and the domain discriminator. In the forward pass, an identity mapping was applied, whereas in the backward pass, the gradient direction was reversed. With the adversarial mechanism, the feature extractor was encouraged to learn domain invariant representations by increasing the domain discriminator loss while reducing the classification error. The overall objective was formed by combining a cross entropy classification loss and a domain adversarial loss. Joint optimization was performed using the Adam optimizer. Offline pretraining was conducted for 30 epochs to learn features that were discriminative for diagnosis and invariant to domain shift.

\subsection{Comparison Experiment}

We compare three different methods. The Baseline refers to the model trained offline using DANN, after which no further online optimization is performed. The Naive strategy conducts online optimization at each time step by treating the model’s own predictions as pseudo-labels and minimizing the cross-entropy loss. All major hyperparameters are kept identical across methods to ensure a fair comparison.
The results are presented in Fig. \ref{Exp} and Table \ref{tab:cumulative_accuracy}.
Under both time-varying operating scenarios, the proposed method achieves the highest diagnostic accuracy across all fault types. In Scenario I, the average cumulative accuracy reaches 92.72\%, outperforming the Baseline (88.86\%) and Naive (77.68\%) methods. In Scenario II, which involves more severe speed-induced variations, the proposed method attains an average accuracy of 80.71\%, exceeding the Baseline and Naive methods by 9.02\% and 35.31\%, respectively.

\begin{table}[htbp]
\centering
\caption{Cumulative Accuracy Comparison Results}
\label{tab:cumulative_accuracy}

  \setlength{\tabcolsep}{3pt} 

\begin{tabular}{lcccccc}

\toprule
\multirow{2}{*}{Fault Type} & \multicolumn{3}{c}{Scenario I} & \multicolumn{3}{c}{Scenario II} \\
\cmidrule(lr){2-4} \cmidrule(lr){5-7}
& Baseline & Naive & Proposed & Baseline & Naive & Proposed \\
\midrule
Teeth crack & 96.09\% & 90.29\% & \textbf{99.46\%} & 71.19\% & 30.12\% & \textbf{79.22\%} \\
Gear wear & 90.55\% & 92.51\% & \textbf{93.26\%} & 72.99\% & 54.93\% & \textbf{80.20\%} \\
Gear pitting & 79.95\% & 50.23\% & \textbf{85.45\%} & 70.88\% & 51.15\% & \textbf{82.70\%} \\
\midrule
Average & 88.86\% & 77.68\% & \textbf{92.72\%} & 71.69\% & 45.40\% & \textbf{80.71\%} \\
\bottomrule
\end{tabular}
\end{table}

Compared with steady operating conditions, transitional conditions impose significantly higher demands on fault diagnosis models. Experimental results show that the Baseline method generally suffers from notable performance fluctuations during condition switching, while the Naive method, which directly updates the model based on unreliable pseudo-labels, is prone to error accumulation in transitional stages, leading to poor performance recovery in subsequent steady phases. In contrast, the proposed method exhibits only a brief performance drop during transitional intervals and rapidly recovers to a stable accuracy level, demonstrating stronger online adaptability and robustness. From the perspective of fault types, the proposed method shows particularly pronounced advantages for weak faults and faults that are highly sensitive to operating condition variations, such as gear pitting and gear wear. These experimental results suggests that the proposed approach not only preserves discriminative representations for prominent fault patterns but also maintains sensitivity to subtle fault characteristics under operating condition perturbations.

Eventually, in Scenario II dominated by speed variations, all methods exhibit lower overall performance than in Scenario I, indicating that speed-induced non-stationarity poses greater challenges than load variations \cite{han2025rethinking}. Nevertheless, even under such severe operating condition shifts, the proposed method maintains relatively stable diagnostic performance, further validating its applicability to strong condition drift and complex online diagnostic scenarios.

\section{Conclusion}\label{section_conclusion}

In this paper, an asymmetric adaptation-based real-time fault diagnosis framework has been proposed to address performance degradation caused by transitional operating conditions in industrial data streams. Domain-invariant representations have been learned in the offline stage. During online inference, asymmetric adaptation has been performed to enable rapid response to operating condition transitions while preserving diagnostic discriminability. Experimental results have demonstrated that the proposed approach has significantly improved robustness and stability under non-stationary operating conditions.

\section*{Acknowledgment}

This work was supported in part by National Natural Science Foundation of China under grants 624B2087, 62525308, 62473223, and 52172323, in part by Beijing Natural Science Foundation under grant L241016. ({\it Corresponding author: Xiao He}.)

\bibliographystyle{ieeetr}
\bibliography{mybibfile}

\end{document}